# New approach for the incoherent and coherent combination of pulsed lasers.

**Hocine.Djellout**

Laboratoire LPCQ, UMMTO, 15000 Tizi-Ouzou
e-mail : hocine.djellout@ummto.dz

*Abstract :*

*This work focuses on the incoherent and coherent combination of pulsed laser beams, building on previous research [1] that addressed the combination of continuous-wave laser beams. For the pulsed combination, we have focused on the temporal evolution of the combined intensity at the image plane. Simulation results show that at the pulse peak, the combined intensity profile is the same, whether in continuous-wave or pulsed mode, for both coherent and incoherent combinations. However, we observe spatial thinning of the combined beams in the case of incoherent pulsed combination when the pulse widths are less than 100 fs. At the front of the pulse, we observe lateral lobes near the focal point, which disappear as we approach the peak of the pulse. Similar to the continuous-wave laser beam combination, our simulations show that in both coherent and incoherent pulsed combinations, the filling factor has no impact on the formation of secondary lobes or on the proportion of energy in the central lobe. Furthermore, with the new configuration, we are no longer constrained by paraxial optics, as the number of lasers that can be combined is arbitrary, and their positions can be chosen freely in space.*

## Introduction:

Achieving a high-quality laser beam with very high powers/energies remains the subject of intense research, due to its applications in various fields such as medicine **[2]**, high-order harmonic generation, which forms the foundation of attosecond science **[3]**, and particle physics **[4-7]** etc. Since its invention in 1985 **[8]**, the chirped pulse amplification technique (CPA) has enabled several orders of magnitude increases in intensity, with intensities on the order of $10^{23}$ W/cm² now being achieved **[9,10]**. However, it appears that the CPA technique is limited by the diameter of Ti:crystals and by transverse amplified spontaneous emission **[11]**, as well as by the size of diffraction gratings. To surpass the 10 PW limit, coherent tiling of four Ti:crystals has been proposed **[12]**, and the coherent combination of laser beams is being explored in high-intensity laser projects such as ELI and XCELS. In our previous work **[1]**, we reported a new configuration and theoretical approach for studying the incoherent and coherent combination of Gaussian laser beams in continuous-wave mode. This approach involves directing all the lasers toward the same focal point, which eliminates the presence of secondary intensity lobes at the focal plane for both coherent and incoherent cases. For the coherent case, the focal spot size depends on the numerical aperture of the entire combined beam system. Furthermore, with the new configuration, we are no longer limited by paraxial optics or by the number of beams that can be combined. However, to achieve high intensities at the focal plane, it is essential to use pulsed lasers. Therefore, in this work, we focus on the study of the combination of spatially and temporally Gaussian laser beams using the same configuration. Simulation results show that at the focal plane and at the peak of the pulse, the combined intensity profile at the focal plane is the same in both continuous and pulsed modes for the coherent case. In contrast, for the incoherent case, with pulses of full width at half maximum (FWHM) smaller than 100 fs, we observe a spatial thinning of the combined intensity. However, the maximum intensity along the on-axis remains the same. At the front of the pulse, we observe lateral lobes near the focal point in both the coherent and incoherent cases.

## Incoherent combination:

The configuration used for the combination of spatially and temporally Gaussian laser beams is the same as that of the previous work **[1]**, and it is shown in Figure 1. The beams to be combined are positioned in the source plane with a minimum waist $w_0$, and the axis of each beam is directed toward the focal point located at the image plane. Each beam, whose center is located at the source plane at coordinates $(x_i, y_j)$, is focused with a lens of focal length $f_{ij}$ in order to optimize the combination. However, since the lasers to be combined are pulsed with FWHM on the order of a few femtoseconds, it is crucial to use off-axis parabolic mirrors to focus the beams and thus avoid dispersion issues.

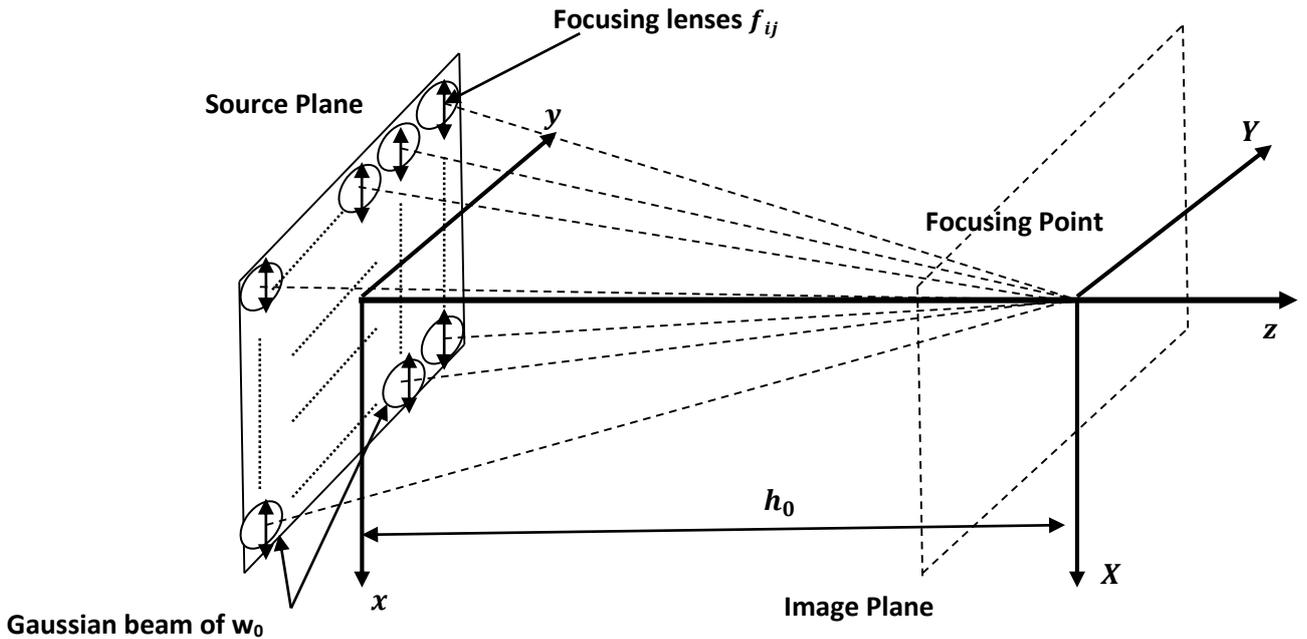

**Fig 1 : New configuration for incoherent and coherent combination of laser beams**

At every point in space, the amplitude of the electric field of a CW Gaussian laser beam is always constant. The same will hold true for the combination of several beams, as the resulting electric field at that point will be the sum of the fields of all beams. However, in the case of a pulsed laser, the amplitude of the electric field at a point in space depends on time. Therefore, for the incoherent combination of pulsed lasers, not only must all beams be directed with high precision toward the focal point, as is done in CW combination, but also the peak intensity of the pulses must reach the focal point simultaneously. To achieve this, it is necessary to use optical path difference controllers in the experimental setup for effective combination, and in the case of coherent combination, a phase-locking system **[13]** is also required to lock the phase of all laser beams at the focal point.

In this study, we chose a Gaussian temporal shape for the laser pulse, the corresponding temporal electric field is described by equation (1).

$$\overrightarrow{E(t)} = \overrightarrow{E_0} exp\left(\frac{-t^2}{2\tau^2}\right) \exp(i\, \omega_0 t) \quad (1)$$

The intensity envelope associated with this field is given by the following expression:

$$I(t) = I_0 \, exp\left(\frac{-t^2}{\tau^2}\right) \quad (2)$$

$\omega_0$ is the carrier frequency, and the FWHM of the pulse is given by $\Delta T = 2\,\tau\,\sqrt{ln2}$.

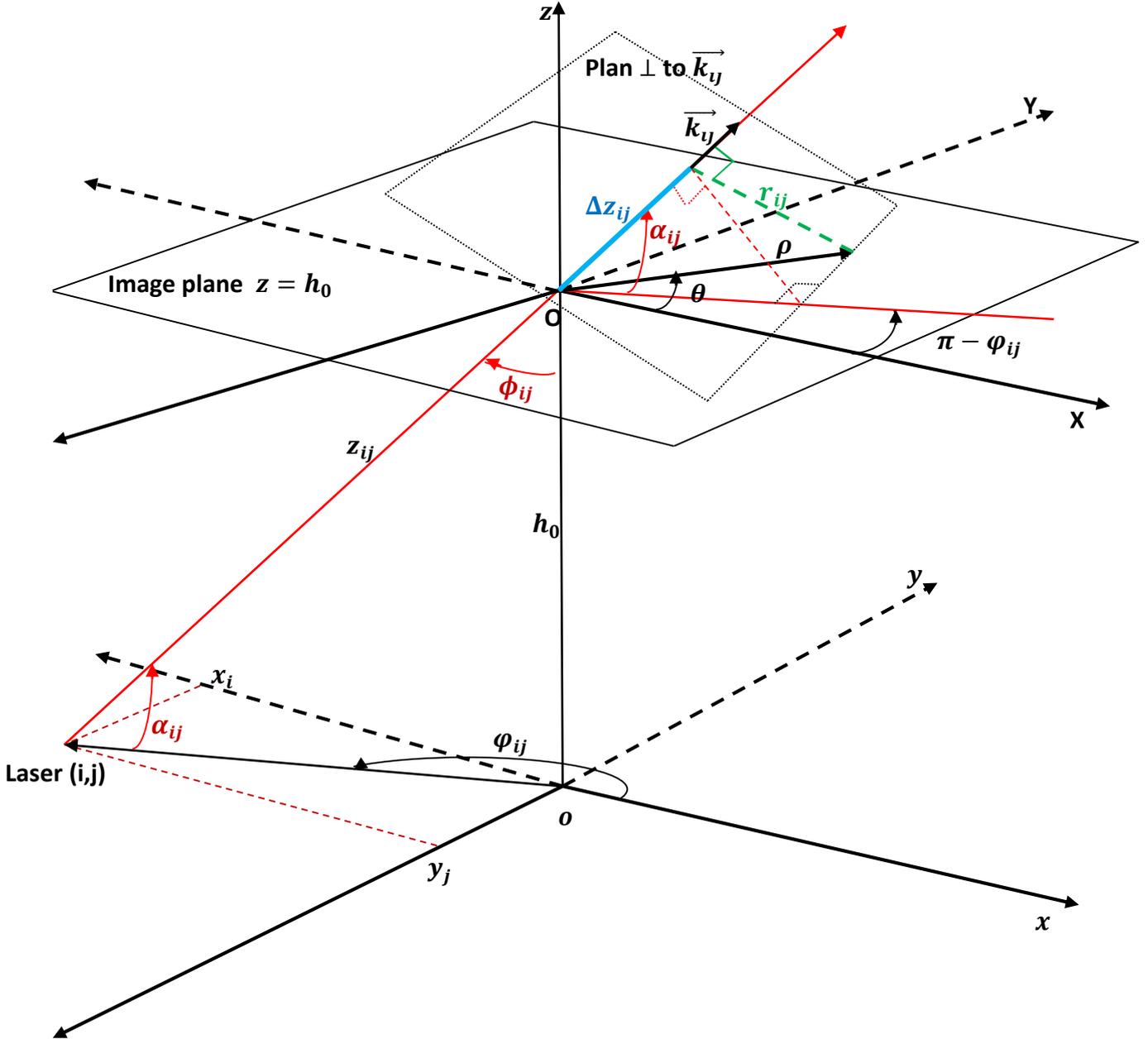

Fig 2: determination of $\Delta z_{ij}$, and $r_{ij}$ at the image plane $z = h_0$

Considering a laser beam with a Gaussian spatial and temporal shape located at the source plane coordinates $(x_i, y_j)$, as shown in Figure 2, our goal is to determine the intensity as a function of time at each point in the image plane. In the case of incoherent combination in CW mode, the intensity of laser (i,j) at the image plane is constant over time and has already been calculated in previous work **[1]**. It is given by:

$$I_{i,j}(\rho,\theta,h_0) = I_0 \frac{w_0^2}{w^2(z_{ij}+\Delta z_{ij})} \exp\left[-2\frac{\rho^2\left(1-\cos^2(\theta-\varphi_{ij}+\pi)\cos^2(\alpha_{ij})\right)}{w^2(z_{ij}+\Delta z_{ij})}\right] \quad (3)$$

With $\Delta z_{ij} = \rho \cos(\theta - \varphi_{ij} + \pi) \cos(\alpha_{ij})$ and $w(z_{ij} + \Delta z_{ij}) = \sqrt{\frac{\lambda}{\pi}} \sqrt{\frac{\left(1-\frac{z_{ij}+\Delta z_{ij}}{z_{ij}}\right)^2 + \left(\frac{z_{ij}+\Delta z_{ij}}{z_0}\right)^2}{\frac{1}{z_0}}}$ is the radius of the focused beam with a focal length $f_{ij} = z_{ij}$ at the distance $z_{ij} + \Delta z_{ij}$.

In the case of incoherent combination in pulsed mode, the intensity at a given point in the image plane depends on time. However, for optimal combination efficiency, the peak of each laser pulse (i,j) from the source plane must reach the focal point, which is the origin of the image plane, simultaneously. For any other point $(\rho, \theta)$ in the image plane, however, the pulse peak will not reach it at the same time as the focal point due to the inclination angle $\alpha_{ij}$ of the laser (i,j) and the spatial extension of the Gaussian beam, as illustrated in Figure 2. Indeed, the pulse peak will reach the point $(\rho, \theta)$ either earlier or later by a time offset given by: $\Delta t_{ij} = \frac{\Delta z_{ij}}{C}$ Where $C$ is the speed of light in a vacuum, and $\Delta z_{ij}$ is the optical path difference for the laser (i,j) relative to the focal point.

To obtain a pulsed intensity, the continuous intensity $I_{i,j}(\rho, \theta)$ should be multiplied by the Gaussian temporal envelope defined by equation (2), taking into account the delay or advance $\Delta t_{ij}$ for a point $(\rho, \theta)$ in the image plane. To perform this multiplication, we assume that there is no spatiotemporal coupling **[14]** for the pulses at the focal plane, so that the spatial and temporal components of the electric field are separable.

Considering a Gaussian pulse shape as shown in Figure 3, let $t$ be the moment when the focal point has an intensity $I(t)$. Since a point $(\rho, \theta)$ in the image plane has a time advance of $\Delta t_{ij} = \frac{\Delta z_{ij}}{C}$ (see Figure 2), its intensity will be equal to $I(t + \Delta t_{ij})$. Conversely, the symmetric point will have an intensity of $I(t - \Delta t_{ij})$. Therefore, the intensity as a function of time for laser (i,j) in the image plane will be equal to:

$$I_{i,j}(\rho,\theta,h_0,t) = I_0 \exp\left(\frac{-\left(t+\frac{\Delta z_{ij}}{C}\right)^2}{\tau^2}\right) \frac{w_0^2}{w^2(z_{ij}+\Delta z_{ij})} \exp\left[-2\frac{\rho^2\left(1-\cos^2(\theta-\varphi_{ij}+\pi)\cos^2(\alpha_{ij})\right)}{w^2(z_{ij}+\Delta z_{ij})}\right] \quad (4)$$

The total time-dependent intensity of the incoherent combination of all laser beams in the image plane will then be equal to:

$$I_T(\rho,\theta,h_0,t) = \sum_{i,j} I_{i,j}(\rho,\theta,h_0,t) \quad (5)$$

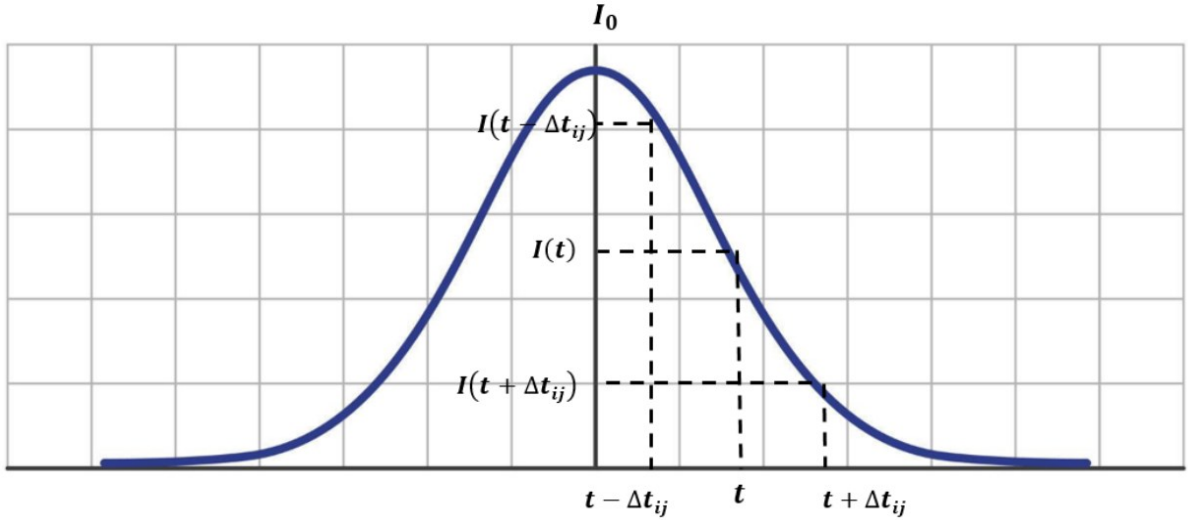

**Fig 3: Advance or delay of $\Delta t_{ij}$ for a point in the image plane for a Gaussian pulse.**

Figure 4 shows the temporal evolution of the combined intensity of 121 incoherent laser beams with Gaussian spatial and temporal profiles, arranged in a square configuration at the source plane. Each laser has a minimum waist of $w_0 = 0.01\,m$, a full width at half maximum of $\Delta T = 10\,fs$, and a central wavelength of $\lambda = 0.8\,\mu m$. All lasers are oriented and focused at the focal point, which is the origin of the image plane located at a distance $h_0 = 1\,m$. Figure 4-a shows the combined intensity at the image plane for $t = 5\Delta T = 50\,fs$ (at the front of the pulse), where we observe four lateral peaks with an intensity of $145.4\,I_0$, located $50.34\,\mu m$ from the focal point. For $t = 3\Delta T = 30\,fs$ (see Figure 4-b), the peak intensity increases to $3.297\times10^4\,I_0$, and the distance from the peaks to the focal point decreases to $32.24\,\mu m$. At $t = 2\Delta T = 20\,fs$ and $t = \Delta T = 10\,fs$ (see Figures 4-c and 4-d), we observe a ring-shaped intensity distribution, with maximum intensities of $3\times10^5\,I_0$ for Figure 4-c and $2.75\times10^6\,I_0$ for Figure 4-d. The distances to the focal point are $22.25\,\mu m$ and $12.8\,\mu m$, respectively. At the peak of the pulse $\Delta T = 0\,fs$ (Figure 4-e), we observe a Gaussian-shaped intensity profile with a waist of $18.9\,\mu m$ and a peak intensity of $1.8\times10^7\,I_0$. Indeed, these observations can be explained by the fact that, at the front of the pulse, the intensity arriving at the periphery of the focal point is greater than that arriving at the focal point itself (as shown in Figure 3), since the beams are tilted. In contrast, at the peak of the pulse, it is the maximum intensity that reaches the focal point, while the periphery of the focal point receives a lower intensity.

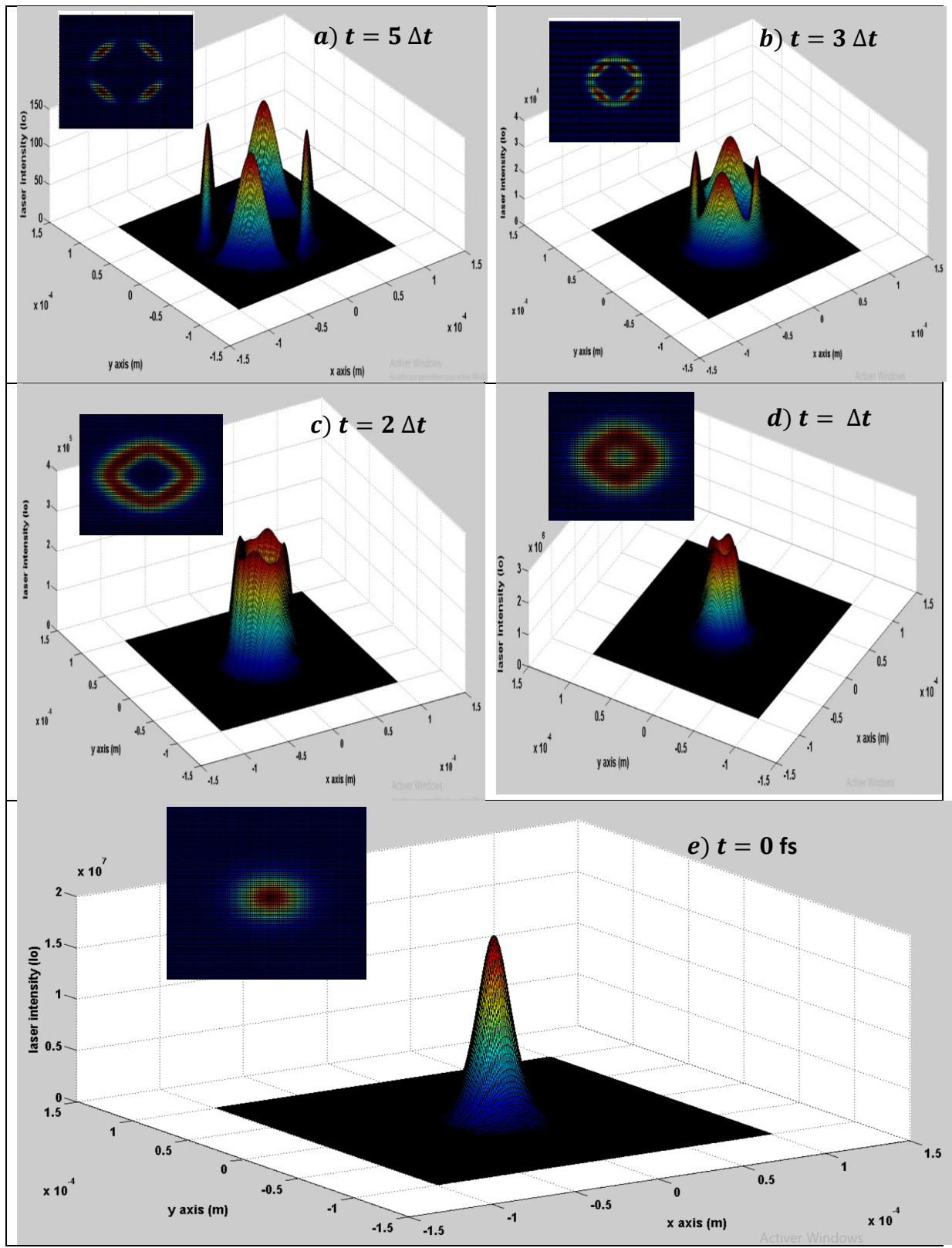

**Fig. 4: Temporal evolution of the incoherent combination of 121 Gaussian lasers with a full-width at half-maximum of 10 fs, focused at 1 m.**

Figure 5 shows the spatial profile of the combined intensity at the pulse peak ($t = 0\,fs$) for different pulse widths. Figure 5a corresponds to $\Delta T = \infty$ (continuous wave mode), while Figure 5b shows results for $\Delta T = 150\,fs$. We observe that both figures are identical, displaying a Gaussian shape with a waist of 26 µm and an on-axis intensity of $1.8 \times 10^7\, I_0$. However, in Figures 5c and 5d, a narrowing of the spatial profile is observed, with waists of 14.5 µm for $\Delta T = 5\,fs$ and 5.1 µm for $\Delta T = 1\,fs$, respectively. These profiles have a spatial shape resembling a Lorentzian, with a square-like base at the bottom of the combined intensity profile. Nonetheless, the on-axis intensity remains the same for all FWHM pulse widths ΔT, maintaining a value of $1.8 \times 10^7\, I_0$. This is explained by the fact that at $t = 0\,fs$, the peak of all laser pulses arrives at the focal point simultaneously, so the combined on-axis intensity equals the sum of the maximum intensities of the pulses. However, for a point located at the periphery of the focal point, the intensity received at $t = 0\,fs$ is not the maximum pulse intensity but a lower value that depends on the position of this point in the image plane. This is related to the spatial extent of the focused Gaussian beam in the focal plane and to the delay or advance of $\Delta t_{ij} = \frac{\Delta z_{ij}}{c} = \frac{\rho\, \cos(\theta - \varphi_{ij} + \pi)\, \cos(\alpha_{ij})}{c}$ as previously explained. Given that the waist of a focused laser beam at the focal plane is approximately 26 µm, this point will have a maximum delay or advance of $\Delta t_{ij} = (26 \times 10^{-6}\,m)/(3 \times 10^8\,m/s) = 86.6\,fs$. For example, a point located 3 µm from the focal point would have a delay or advance of $\Delta t_{ij} = 10\,fs$. The further the point is from the focal point, the greater the delay or advance, resulting in a lower intensity compared to the continuous wave case, which explains the narrowing of the spatial profile of the combined intensity, as shown in Figure 6.

For FWHM pulse widths greater than $\Delta T = 100\,fs$, the intensity at this point will be practically the same as in the continuous case. In contrast, for FWHM pulse widths less than $\Delta T = 100\,fs$, the intensity at the point will be lower than in the continuous case. The minimum combined waist is obtained for FWHM pulse widths of $\Delta T = 1\,fs$, with a value of 5.1 µm. It is possible to further decrease the combined waist by reducing ΔT, but it cannot go beyond the diffraction limit, as for ΔT below a femtosecond, the carrier wavelength will differ, becoming shorter.

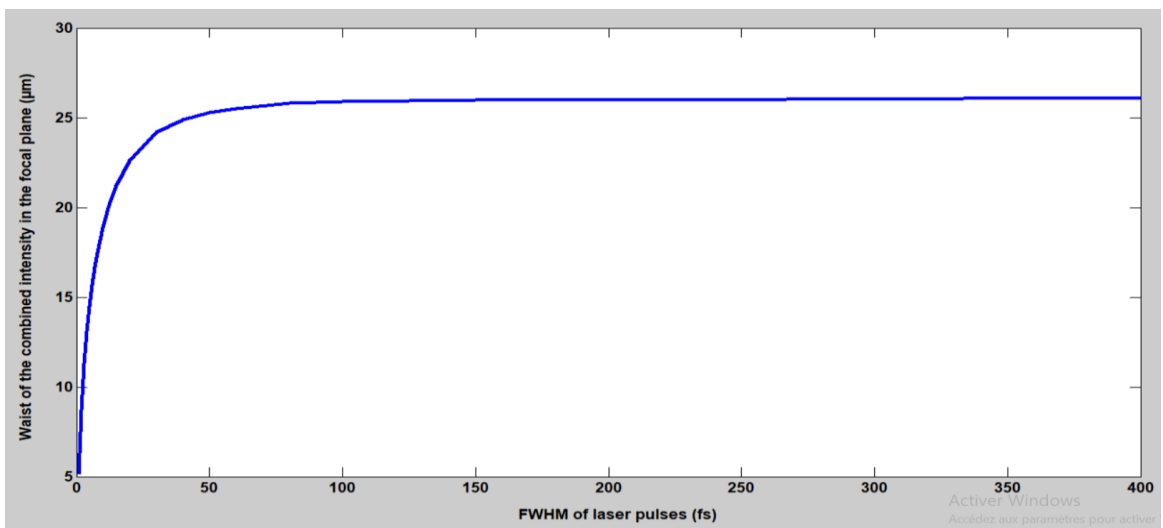

**Fig 6: Waist of the combined intensity in the focal plane as a function of the FWHM ΔT, of the laser pulses.**

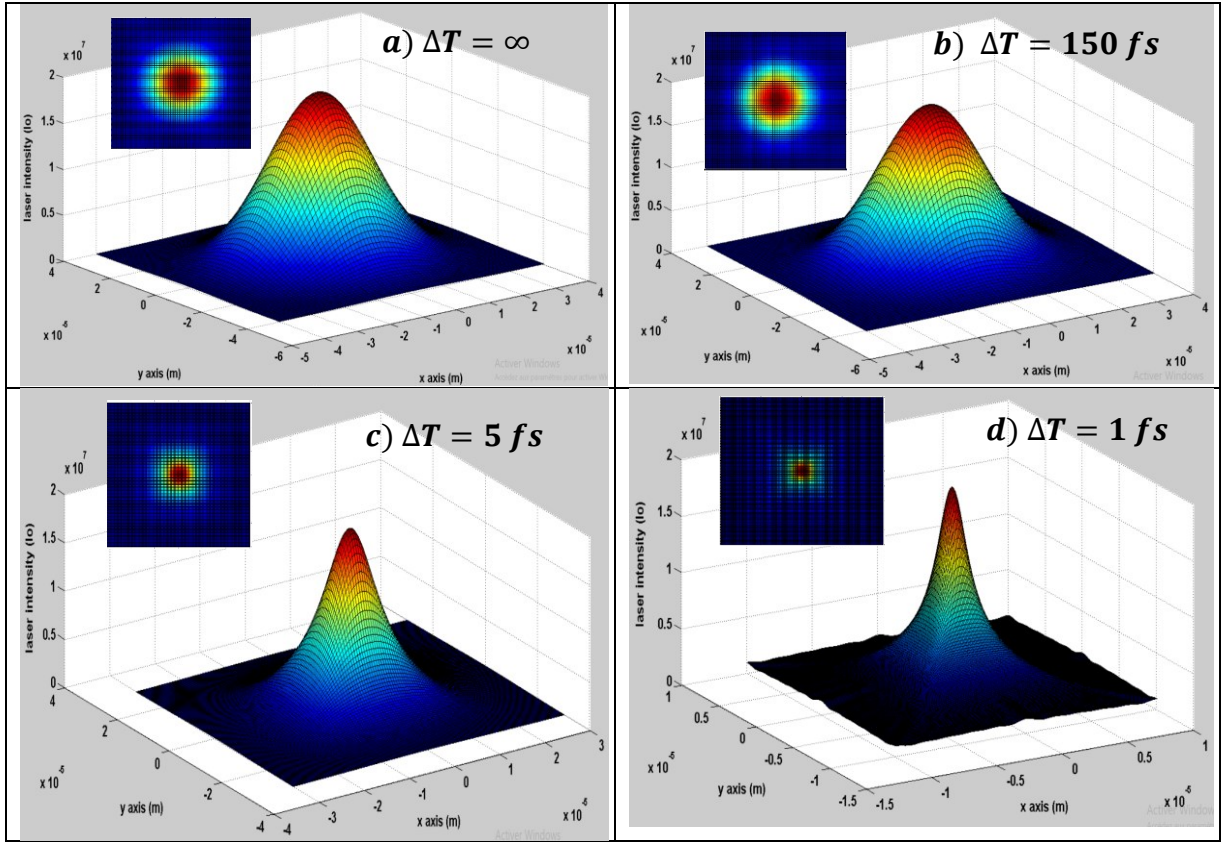

**Fig 5: Spatial profile of the combined intensity for different FWHM of the laser pulses. a) ΔT=∞, b) ΔT=150 fs, c) ΔT=5 fs, d) ΔT=1 fs**

## Coherent Combination:

In the case of incoherent combination, the peaks of the laser pulses must reach the focal point simultaneously. However, for coherent combination, all laser beams must also be in phase at the focal point; this requires phase locking at the focal point. The electric field of the coherent combination in continuous mode at the focal plane is given by the following equation **[1]**:

$$\vec{E_T}(\rho, \theta, h_0) = \sum_{i,j} \sqrt{I_0} \left( \frac{h_0}{\sqrt{x_i^2 + h_0^2}} \vec{i} + \frac{x_i}{\sqrt{x_i^2 + h_0^2}} \vec{e_z} \right) \frac{w_0}{w(z_{ij} + \Delta z_{ij})} \exp\left[ -\frac{r_{ij}^2}{w^2(z_{ij} + \Delta z_{ij})} \right] \exp\left( i2\pi \rho \, \cos(\theta - \varphi_{ij} + \pi) \, \frac{\sin(\phi_{ij})}{\lambda} \right) \quad (6)$$

The exponential term $\left( 2\pi \rho \, \cos(\theta - \varphi_{ij} + \pi) \, \frac{\sin(\phi_{ij})}{\lambda} \right)$ represents the phase shift of the field from laser $(i,j)$ at a point in the image plane $(\rho, \theta)$ relative to the focal point. This phase shift depends on the wavelength, and unlike in the monochromatic case, a femtosecond laser pulse consists of a broad spectrum. Therefore, a rigorous calculation of coherent pulsed combination must account for the interference of all spectral components of the pulses.

At the focal point, all spectral components of all lasers $(i,j)$ are in phase. However, as we move away from the focal point, they no longer remain in phase due to their different wavelengths. Consequently, a partially coherent superposition of the focal spots for all spectral components of the pulses is expected. However, since $\Delta\lambda/\lambda \approx 0.12$ for a 10 $fs$ Gaussian pulse (Fourier-limited) with an 800 nm carrier, and this ratio is even lower for longer pulses, and given that a large portion of the pulse energy falls within the spectral bandwidth $\Delta\lambda$, the slowly varying envelope approximation, where the phase term of the pulse is represented only by the carrier wavelength, may serve as a reasonable first approximation for studying the coherent combination of pulsed lasers.

Similar to incoherent pulsed combination, one must also consider the delay or advance $\Delta t_{ij}$ for a point $(\rho, \theta)$ in the image plane. Thus, the electric field of the coherent combination in pulsed mode at the focal plane can be expressed as:

$$\vec{E_T}(\rho,\theta,h_0,t) = \sum_{i,j} \sqrt{I_0 exp\left(\frac{-\left(t+\frac{\Delta z_{ij}}{C}\right)^2}{\tau^2}\right)} \left(\frac{h_0}{\sqrt{x_i^2+h_0^2}}\vec{i} + \frac{x_i}{\sqrt{x_i^2+h_0^2}}\vec{e_z}\right) \frac{w_0}{w(z_{ij}+\Delta z_{ij})} exp\left[-\frac{r_{ij}^2}{w^2(z_{ij}+\Delta z_{ij})}\right] exp\left(i2\pi\rho \cos(\theta-\varphi_{ij}+\pi)\frac{\sin(\phi_{ij})}{\lambda}\right) (7)$$

The total intensity in the image plane is then given by:

$$I_T(\rho,\theta,h_0,t) = \left|\vec{E_T}\vec{E_T}^*\right| \quad (8)$$

We chose to combine around a hundred lasers because it has been reported in the literature that several research teams have coherently combined multiple tens of lasers in a "tiled aperture" configuration, both in continuous mode **[15-17]** and pulsed mode **[18]**. Figure 7 shows the temporal evolution of the combined intensity of 121 coherent laser beams with Gaussian spatial and temporal profiles. Similar to incoherent combination, we observe side peaks near the focal point, as shown in Figures 7a, 7b, and 7c. However, unlike incoherent combination, there are more peaks due to interference between the different beams.

At $t = \Delta T$ (Figure 7d), we observe five high-intensity peaks: four side peaks and one central peak, followed by several low-intensity lobes aligned along the x and y axes. This pattern resembles the Fourier transform of a square, as the lasers in the source plane are arranged in a square formation. The central peak has an intensity of $1.347 \times 10^8 I_0$. and a waist of approximately 1.35 µm. At $t = 0\ fs$, at the peak of the pulse (Figure 7e), there is a central Gaussian-shaped peak with an intensity of $2.154 \times 10^9 I_0$ and a waist of around 1.35 µm, along with secondary lobes along the x and y axes, also representing the Fourier transform of a square.

Unlike incoherent combination, where we observe a spatial narrowing of the combined intensity for pulses shorter than 100 $fs$, in the case of coherent combination, we observe almost no difference in the spatial profile of the combined intensity, whether in continuous mode ($\Delta T = \infty$) or pulsed mode for full widths at half maximum greater than $\Delta T = 3\ fs$ (see Figure 8). However, for $\Delta T < 3\ fs$, we observe a slight narrowing, even for a lower numerical aperture and therefore a larger focal spot. Simulations show that the focal spot is no longer a Fourier transform of a square, and we observe a slight narrowing of the central lobe, where the full width at half maximum of the central lobe decreases from 1.7 µm for a 10 $fs$ pulse to 1.5 µm for a 1 $fs$ pulse. This result does not allow us to claim that we can go beyond the diffraction limit, as the slowly varying envelope approximation is already at its limits for $\Delta T = 10\ fs$. For sub-femtosecond pulses, the carrier wavelength falls within the UV range, Thus, a more rigorous study that takes into account the interference of all spectral components of all pulses would be interesting to see if it is possible to go beyond the diffraction limit, which is, in essence, an uncertainty principle **[1].**

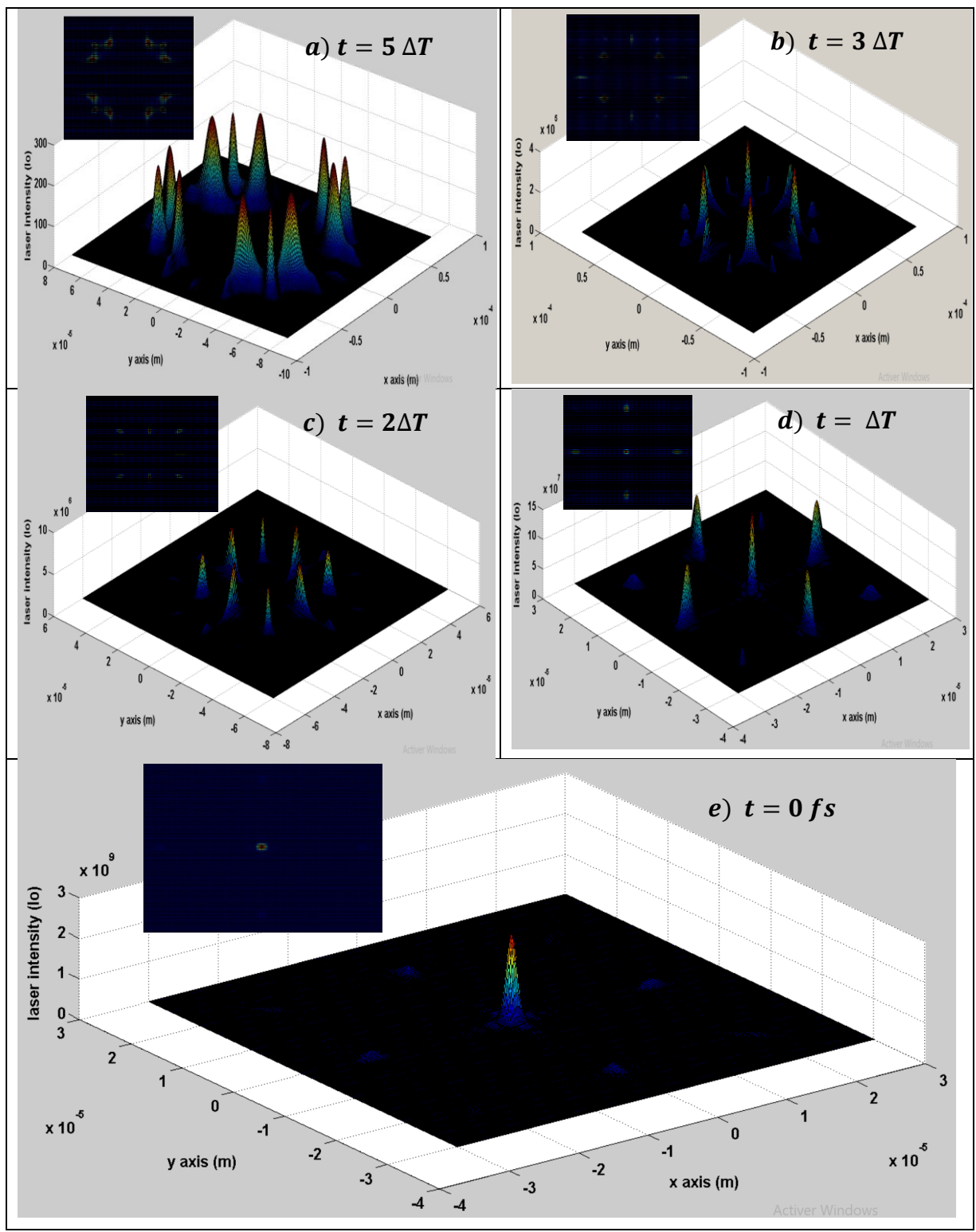

**Fig 7: Temporal evolution of the coherent combination of 121 Gaussian lasers with a FWHM of 10 fs, focused at 1 m.**

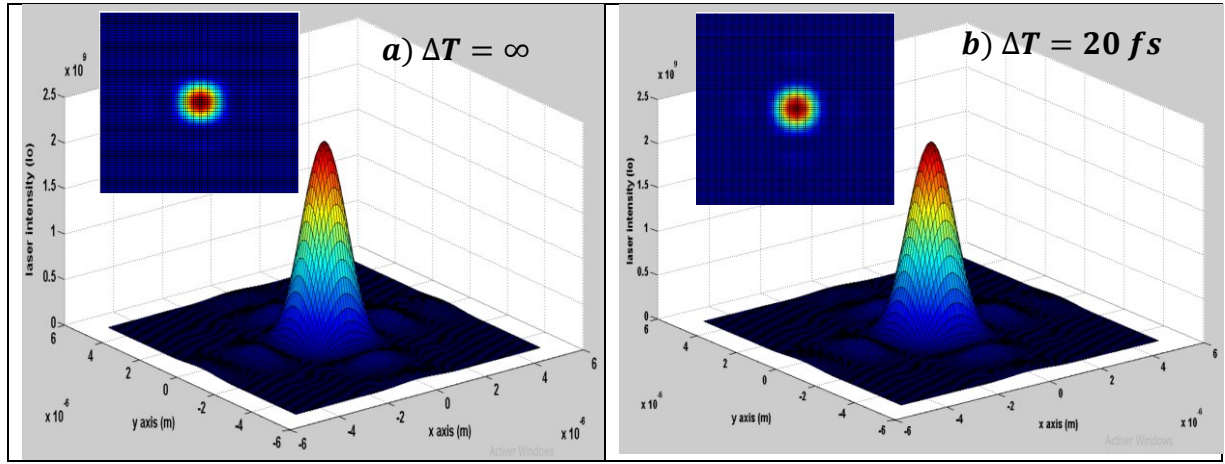

**Fig 8: Spatial profile of the coherent combination for different laser pulses a) $\Delta T = \infty$ (CW), b) $\Delta T = 20\ fs$.**

**Effect of the filling factor on the spatial profile of the combined intensity:**

The literature reports that optimizing the filling factor (minimum spatial gap between beams at the source plane) can optimize the spatial profile of the combined intensity and the proportion of energy contained in the central lobe in the far field **[13,19-20]**. In this work, we demonstrate that, with the new configuration, the filling factor has no effect on either the spatial profile of the combined intensity or the proportion of energy carried by the central lobe, whether in coherent or incoherent combination, in continuous or pulsed mode.

For this, in our simulations, we varied the distance between the centers of adjacent beams at the source plane from $0.04\ m$ to $0.07\ m$ to observe the effect of reducing the filling factor. As shown in Figure 9, obtained with 121 pulsed lasers with a FWHM of $20\ fs$, We observe that there is no difference in the spatial profile of the incoherent combination in the pulsed mode for an adjacent beam spacing of 0.04 m (Figure 9-b) and 0.07 m (Figure 9-a). The spatial profile remains Gaussian, but we note a slight difference in intensity and waist radius: $1.809 \times 10^7\ I_0$ and 22.5 μm for a 0.04 m laser distance, and $1.705 \times 10^7\ I_0$ and 20.5 μm for a 0.07 $m$ distance. This is because the waist of the focused lasers at the image plane varies negligibly between the two configurations ($0.04\ m$ and $0.07\ m$).

We observe the same for the coherent combination in pulsed mode, as shown in Figure 9-c ($0.07\ m$) and Figure 9-d ($0.04\ m$), where there is always a central Gaussian-shaped peak with secondary lobes along the x and y axes, which is the Fourier transform of a square. However, we observe a reduction in waist from 1.35 μm for a $0.04\ m$ distance between lasers to 0.8 μm for a $0.07\ m$ distance. This is because increasing the distance between lasers at the source plane increases the system's numerical aperture, resulting in a reduction in focal spot size. We observe the same effect for the continuous mode combination.

Thus, we understand that with this new configuration, the filling factor has no effect on either the spatial shape of the combined intensity or the energy proportion carried by the central lobe. Furthermore, this new configuration is equivalent to a lens of any size, and we are no longer limited by paraxial optics, as all that matters is synchronizing, phasing, and directing all beams to the focal point. Even the spatial positioning of the beams is irrelevant, as per the equation $\Delta\rho\, ON \geq \lambda/2$ derived previously [1], meaning the critical factor is the configuration's numerical aperture ($ON$) relative to the focal point.

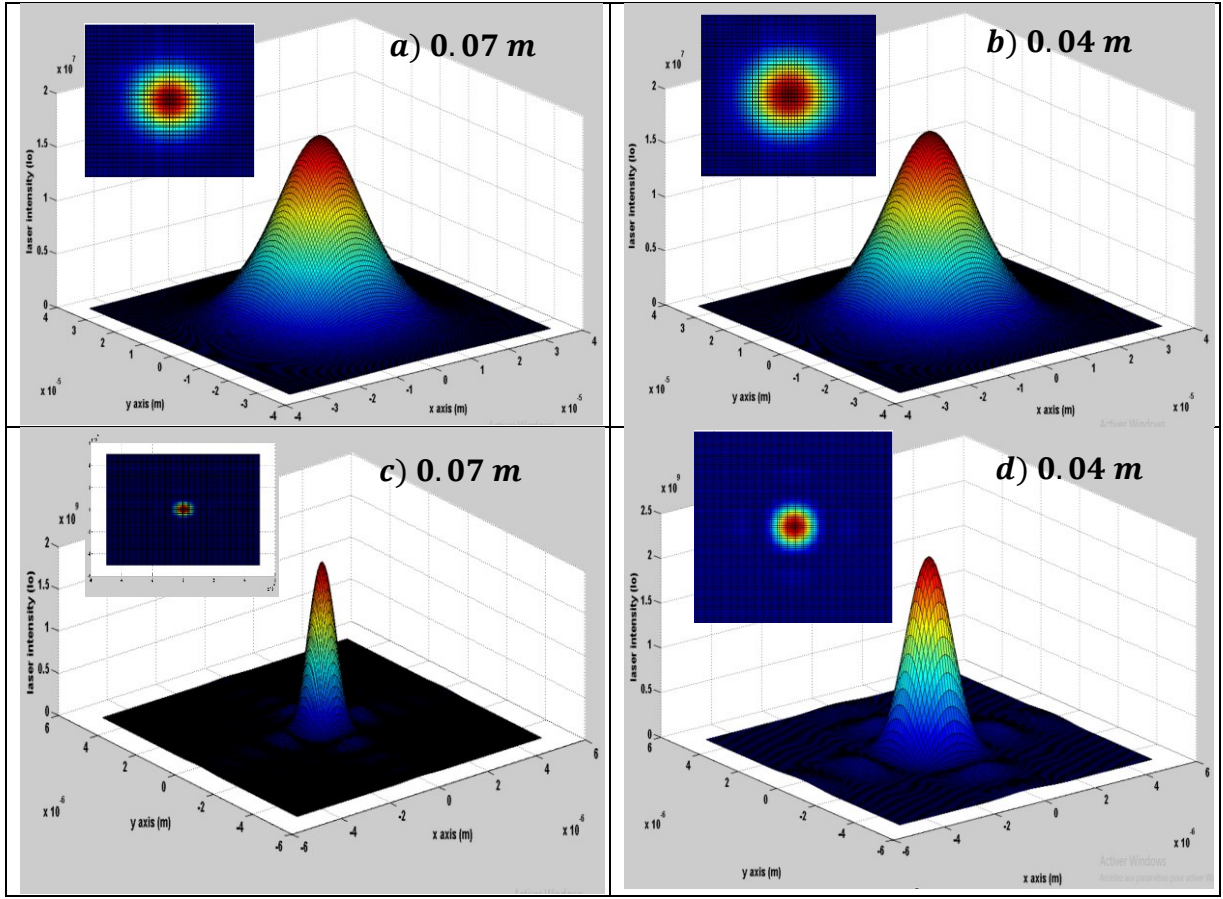

**Fig. 9: Effect of the fill factor (distance between two adjacent lasers at the source plane) on the spatial profile of the combined intensity in pulsed mode for the incoherent case (a and b) and the coherent case (c and d).**

To achieve ultra-intense peaks, coherent combination of multiple pulsed laser beams based on Chirped Pulse Amplification (CPA) would be advantageous. Recently, a team demonstrated the feasibility of coherently combining four femtosecond laser beams based on CPA [21]. However, to reach ultra-high intensities, combining several beams is desirable. To this end, we simulated the combination of 121 pulsed laser beams using our new configuration, with a beam waist of $w_0 = 75\ mm$ before focusing, and a distance of $300\ mm$ between the centers of adjacent beams. All beams are arranged in a square layout covering an area of approximately $9\ m^2$ at the source plane, and are focused at the focal point of the image plane, located $1.5\ m$ from the source plane. The FWHM of the pulses is $20\ fs$, and each laser beam has a power of $40\ TW$.

Figure 10 shows the result of this coherent combination in pulsed mode. The spatial profile of the combined intensity displays a central lobe with a Gaussian shape and weaker side lobes aligned along the x- and y-axes. The waist of the central lobe is 0.37 μm, and the on-axis intensity at the focal point is $1.431 \times 10^{12}\ I_0$, corresponding to an on-axis intensity of approximately $0.65 \times 10^{24}$ W/cm$^2$. This is an intensity beyond the capabilities of current ultra-intense lasers. Moreover, the simulated configuration is not yet optimized, so we believe that with this new configuration, if we can phase-locking a large number of high power pulsed lasers at the focal point, it may be possible to achieve even higher intensities.

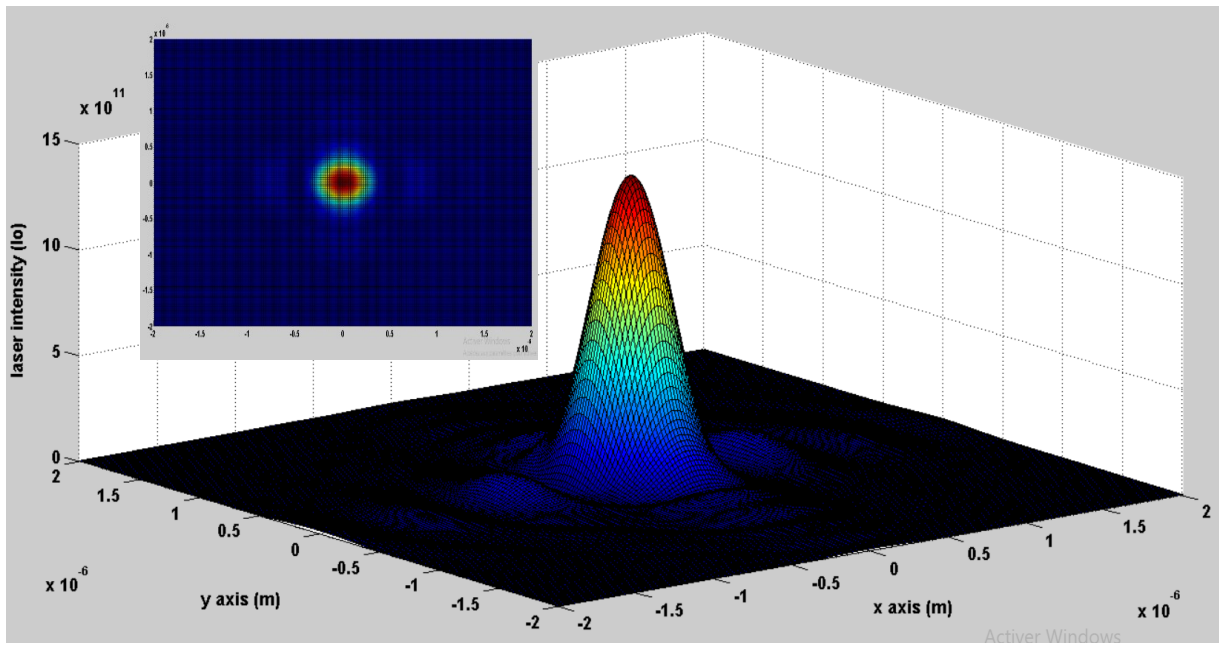

**Fig. 10: Coherent pulsed combination of 121 lasers of 40 TW.**

## Conclusion

In this work, we extended the study of coherent and incoherent combination from continuous to pulsed mode, allowing us to determine the temporal evolution of the combined intensity at the image plane. At the peak of the pulse, we observed that the spatial profile of the combined intensity is the same as in the continuous mode. However, we noted a narrowing of the central lobe in the case of incoherent combination for pulses shorter than $100\ fs$. We also showed that the fill factor has no influence on either the shape or the energy of the central lobe of the combined intensity. We believe that optimizing this new configuration will enable record-breaking intensities if a large number of high power laser beams can be phase-locked at the focal point.